\documentclass[
  ,draft            
  ]
  {aipproc}

\layoutstyle{6x9}

\begin{document}

\title{Black Hole States: Accretion and Jet Ejection}


\author{Belloni T.}{
  address={INAF-Osservatorio Astronomico di Brera, Via E. Bianchi 46,
  	I-23807 Merate, Italy}
}

\begin{abstract}
The complex spectral and timing properties of the high-energy
emission from the accretion flow in black-hole binaries, 
together with their strong connection to the 
ejection of powerful relativistic jets from the system, can be simplified
and reduced to four basic states: hard, hard-intermediate,
soft-intermediate and soft. Unlike other classifications, these states
are based on the presence of sharp state transitions. I summarize this
classification and discuss the relation between these states
and the physical components contributing to the emitted flux.

\end{abstract}

\maketitle


\section{Introduction}

Since the launch of RossiXTE, our knowledge of the high-energy
emission of Black Hole Binaries (BHB) has increased enormously,
leading to a new, complex picture which is difficult to 
interpret. At the same time, a clear connection between X-ray and
radio properties has been found (see Fender 2005). In the following, I 
present briefly the state paradigm that is now emerging, based
on a large wealth of RossiXTE data from bright transient sources
and its connection with jet ejection
(see Belloni et al. 2005; Homan \& Belloni 2005; Fender, Belloni
\& Gallo 2004). I discuss this
in general terms, in the attempt to simplify the picture as much as possible. 

\section{Black Hole States and Jet Ejection}

The results of detailed timing and color/spectral analysis of the RossiXTE data
of bright BHBs have evidenced a very wide range of phenomena which are 
difficult to categorize. Nevertheless, it is useful to identify distinct
states. Based on the variability and spectral behavior and the transitions
observed in different energy bands, we consider the following states in addition
to a quiescent state (see Homan \& Belloni 2005; Belloni et al. 2005; 
Casella et al. 2004,2005 for the description of the different QPO types):

\begin{itemize}

\item Low/Hard State (LS): this state is associated to relatively low
values of the accretion rate, i.e. lower than in the other bright states. The
energy spectrum is hard and the fast time variability is dominated by a 
strong ($\sim$30\% fractional rms) band-limited noise. Sometimes, low
frequency QPOs are observed. The characteristic frequencies detected in the
power spectra follow broad-range correlations (see Belloni, Psaltis \& van der
Klis 2002). In this state, flat-spectrum radio emission is observed, 
associated to compact jet ejection (see Gallo, Fender \& Pooley 2003; 
Fender, Belloni \& Gallo 2004). 

\item Hard Intermediate State (HIMS): in this state, the energy spectrum is 
softer than in the LS, with evidence for a soft thermal disk component. The
power spectra feature band-limited noise with characteristic frequency 
higher than the LS and usually a rather strong 0.1-15 Hz type-C QPO
(see e.g. Casella et al. 2005). The
frequencies of the main components detected in the power spectra
extend the broad correlations mentioned for the LS. 
The radio emission shows a slightly steeper spectrum (Fender, Belloni 
\& Gallo 2004). Just before the transition to the SIMS (see below), 
Fender, Belloni \& Gallo (2004) suggested that the
jet velocity increases rapidly, giving origin to a fast relativistic jet.

\item Soft Intermediate State (SIMS): here the energy spectrum is systematically
softer than the HIMS. The disk component dominates the flux. No strong 
band-limited noise is observed, but transient type-A and type-B QPOs, the 
frequency of which spans only a limited range. No core radio emission is
detected.

\item High/Soft State (HS): the energy spectrum is very soft and 
strongly dominated by a thermal disk component. Only weak power-law noise is observed
in the power spectrum. No core radio emission is detected
(see Fender et al. 1999; Fender 2005).

\end{itemize}

\begin{figure}
  \includegraphics[height=.4\textheight]{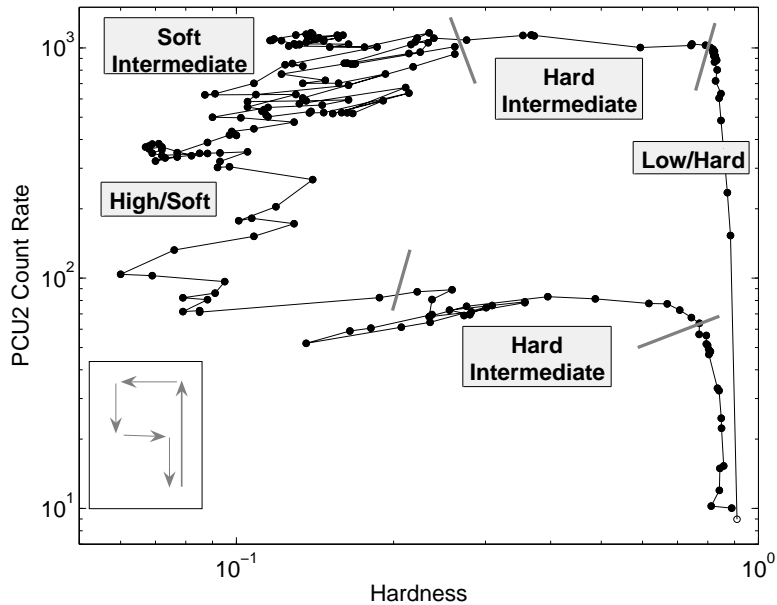}
  \caption{Hardness-Intensity diagram of the 2002/2003 outburst of
  	GX 339-4 as observed by the RXTE PCA. The gray lines mark
	the state transitions described in the text. The inset on the 
	lower right shows the general time evolution of the outburst
	along the 'q'-shaped pattern.
	From Belloni et al. (2005).}
\end{figure}

This simplified classification needs to be supported by a picture of the time evolution
in a transient source. The states described above are defined also in 
terms of their transitions, which need to be taken into account. A sketch
of the evolution of the 2002/2003 outburst of GX 339-4 
is ideal to show these transitions (see Homan \& Belloni 2005; 
Belloni et al. 2005).
Figure 1 shows the outburst in a Hardness-Intensity Diagram (HID): the x-axis
shows the X-ray hardness and the y-axis the detected count rate (from 
Belloni et al. 2005). The outburst can be described in the following way.

\begin{itemize}

\item The system starts its outburst as a weak 
hard source (the bottom right in the HID): the first detection with the
RXTE/PCA (a factor of $\sim$10 lower than the first point in Fig. 1), 
indicates an X-ray flux a factor of $\sim$10 higher than in 
quiescence (Homan et al. 2005). The flux increases with time, while the
X-ray colors indicate that the spectrum is still hard and only mildly softening: 
the source moves upward in the right branch. During this period, the infrared
flux was seen to correlate strongly with the X rays, indicating that both
components are connected, possibly originating from a jet (Homan et al. 2005).
This branch corresponds to the LS, when flat-spectrum radio emission is
observed  (see Fender et al. 1999; Fender, Belloni \& Gallo 2004). The 
characteristic frequencies of the different noise components
in the power spectrum increase with time. 
The energy spectrum is roughly described by 
a cutoff-power law. 

\item At the top of the right branch, GX 339-4 moves left and enters a
	horizontal branch, corresponding to the HIMS. The transition is
	shown by the dotted line. The precise position of this
	line corresponds to a marked change
	in the IR/X correlation (Homan et al. 2005). The drifting to the
	left is the result of two causes: the appearing of a thermal 
	disk component and the steepening of the power-law component. In GRS 1915+105,
	recent Integral/RXTE observations (Rodriguez et al. 2004) 
	indicate that in this state two hard components are at work (see
	also Zdziarski et al. (2001). In the power spectra, the characteristic
	frequencies continue to increase so that the transition in the
	timing domain appears to be rather smooth.
	As GX 339-4 moves left in the HID, the radio spectrum steepens
	slightly and as the source approaches
	the `jet line' (see below) the velocity of the jet outflow increases
	(Fender, Belloni \& Gallo 2004).

\item When the source, moving left, passes the second dotted line, a very sharp
	transition is observed. The variation of X-ray hardness is small,
	indicating that the energy spectrum does not change by a large amount. 
	However, the power spectrum changes abruptly: the strong
	band-limited noise and type-C QPO disappear and a sharp type-B
	QPO appears. The source has entered the SIMS.
	After this transition, the source remains for a long time in the upper
left quadrant. The spectrum is soft and dominated by the thermal disk component,
and the power spectrum shows complex and variable features. Transient type-B
QPOs are observed, but always when the hardness is comparable to that of the
first detection. The sharp transition, observed in other systems as well
(see Homan \& Belloni 2005; Casella et al. 2004) corresponds to the crossing of
the `jet line' (in the terminology of Fender, Belloni \& Gallo 2004). 
It is in correspondence to the crossing of the jet line
	that the strong radio flare is observed (Gallo et al. 2004) and a fast
	relativistic jet is launched.  This is also observed in other systems
	such as GRS 1915+105, XTE J1859+226 and XTE J1550-564. 
	
\item After entering the SIMS, the source remains for a long time (about
five months) in the upper
left quadrant, moving in a complex fashion (see Fig. 1). 
The spectrum is soft and dominated by the thermal disk component,
and the power spectrum shows very complex and varying features. Transient type-B
QPOs are observed, as well as other weak features. However, no strong
band-limited noise component is observed. The properties of the SIMS are
complex, and in other systems such as XTE J1859+226 fast transitions to and from
the HIMS (i.e. between type-A/B and type-C QPOs) are observed (Casella et al.
2004), corresponding to the crossing of the jet line. For that source, there is
evidence that subsequent radio flares are observed in correspondence of these
transitions (see Casella et al., this volume). During the SIMS, observed radio
emission corresponds only to the previous jet ejections 
(Fender, Belloni \& Gallo 2004).

\item Once the count rate of GX 339-4 goes below a  few hundred counts per second
per PCU, it enters the HS. The spectrum remains soft, the flux decreases with
time, and the power spectrum only features a weak power-law component. Even
integrating all the HS observations, no other timing component appears. No 
radio emission is observed in the HS (see Fender 2005).

\item Below $\sim$100 cts/s/PCU, the source hardens again and enters the HIMS
(third dotted line in Fig. 1). Notice that for three observations the source is
found back in the range of hardness corresponding to the HS, and indeed HS
properties are observed at those times in the power spectrum. No clear instance
of SIMS is seen, although one observation at hardness similar to the bright SIMS
points does show a clear 1 Hz QPO which could be a low-frequency version of the
type-B ones, which were observed around 6-7 Hz at high flux. 

\item As GX 339-4 continues to harden, it moves back to the LS with a 
smooth transition. The points in the HID
reach almost the same position where they started the outburst.

\end{itemize}

\begin{figure}
  \includegraphics[height=.3\textheight]{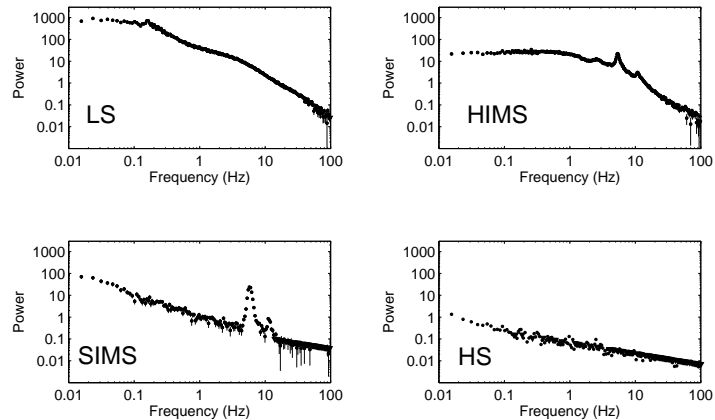}
  \caption{Examples for power spectra of GX 339-4 corresponding to the four
  states described in the text (adapted from Homan \& Belloni 2005).}
\end{figure}

This picture, albeit simplified, provides a useful
framework within which one can work on a modeling of the properties of the
accretion flow and the ejection of collimated jets. This classification
differs from that proposed by McClintock \& Remillard (2005), being based
on the presence of sharp state transitions rather than 
on a characterization in terms of spectral and timing parameters.

\section{Physical states}

In order to understand the scheme outlined above, we should
consider the main components that are observed from the system.

\subsection{Spectral components}
There is amounting indication that three main spectral continuum 
components contribute
to the observed high-energy flux (I deliberately ignore additional
components such as emission lines and Compton reflection). 
The first is the thermal thin disk, 
observed at energies below 10 keV. This component is probably present in the
LS, but its temperature is in most cases too low to be observable (see 
e.g. the case of XTE J1118+480; Frontera et al. 2001; McClintock et al. 2001).
In the HIMS it contributes a small (varying) degree to the X-ray flux, while it is
dominant it in the SIMS and the HS. This component appears to be rather
quiet and does not contribute much to the observed fast variability. 
There is evidence that the innermost radius of this thin disk, as evaluated
from the energy spectrum, decreases from large values in the LS to smaller
values in the HS, although absolute measurements of this radius are
plagued by a number of problems (see e.g. Merloni, Fabian \& Ross 2000).

The second component is much harder and is usually fitted with a thermal
Comptonization from $\sim$100 keV electrons or with its approximation as 
a power-law with a 
high-energy cutoff. This is the component that is observed in the LS and
which is clearly associated to the strong band-limited noise typical of this
state. In recent years, the possibility that this component originates
directly from the jet has been proposed and fits to the broad-band
spectra of LS sources have been attempted (see e.g. Markoff et al. 2003).
It is not observed in the HS.

The third component is modeled with a steep power law, observed in the HS
and which has been reported also for a variety of spectra the state
attribution of which is uncertain (see Grove et al. 1998) but in some cases
probably corresponding to the HIMS. This component was
detected in GRS 1915+105, which is always observed in the HIMS (Zdziarski et
al. 2001) and attributed to non-thermal Comptonization.
For the SIMS, there is some evidence that no high-energy cutoff is observed
with HEXTE, indicating that this component is present (Homan et al., in prep.).
Interestingly, Integral/RXTE observations of GRS 1915+105 reported
by Rodriguez et al. (2004) show that the combined timing/spectral properties can
be explained with the simultaneous presence of both hard components, one with a
high-energy cutoff and one without. Therefore, the scenario that is emerging 
for the high-energy component is
one with a thermal component in the LS and a non-thermal one in the HS, while in
the HIMS, which is intermediate between these two, both components are present.
It is therefore possible that in the top branch in Fig. 1 (HIMS) the 
thermal hard component decreases and the non-thermal hard component appears, 
although it
is not clear how gradual these changes are. After the jet line is crossed,
however, the thermal hard component disappears.

\subsection{Radio emission and jets}

Fender, Belloni \& Gallo (2004) proposed a unified scheme in which the compact
flat-spectrum jet emission and the fast relativistic jet component originate
from the same outflow, which is accelerated to high Lorentz factors as the
source approaches the jet line. Indeed, core radio emission is observed only 
in the LS (down to quiescence level, see Gallo, Fender \& Hynes 2004) 
and HIMS.  The remaining two states, SIMS and HS, do not show radio
emission.

\subsection{Noise and QPO components}

As shown above, the properties of the fast time variability are complex,
but it is useful to examine their behavior in relation
to the states described above. Two states, LS and HIMS, show strong band-limited
noise and type-C QPOs; the frequencies of these power-spectrum components follow 
correlations that suggest there is a one-to-one correspondence between 
components in the two states (see Belloni, Psaltis \& van der Klis 2002). 
Interestingly, these frequencies vary over 
a large range in the states where large variations in the inferred inner disk
radius are observed. Their energy spectrum is hard (see e.g. Rodriguez
et al. 2004; Casella et al. 2004). 
The two soft states, HS and SIMS, while displaying a number of transient
feature such as type-A/B QPOs and other weak components difficult to 
classify (see Casella et al. 2004,2005; Belloni et al. 2005), never show strong
band-limited noise components such as those described above. The QPOs
in these states, in particular type-B QPOs, vary over a much smaller range of
frequencies compared to the type-C ones. Their spectrum is also hard
(Casella et al. 2004).
The few instances of high frequency QPOs in BHBs were all observed in 
the SIMS (see e.g. Morgan, Remillard \& Greiner 1997; 
Homan et al. 2001,2003; Cui et al. 2000; Remillard et al. 1999).

\section{Discussion}

Above, we discussed the presence of four states of BHBs, which are identified
in terms of the presence/absence of three spectral components, two or more
power-spectral components, and a radio-emitting jet with varying Lorentz
factor.

Before the 90's, when only relatively sparse data were available, two BHB states
were clearly defined: a low/hard state and a high/soft state.
The general picture that can be drawn from the properties discussed above
is now the following. In the hard
state, the source is jet-dominated, in the sense that the power in 
the jet is probably larger than that in the accretion (see Fender, Gallo \& Jonker
2003). The dominant component in the X-ray range is the thermal hard component,
which is possibly associated to the jet itself. 
The geometrically thin, optically thick, accretion
disk is very soft and has a varying inner radius, so that its contribution to 
the X-ray emission changes. This state is characterized by strong
band-limited noise and type-C QPOs, whose characteristic frequencies show
clear correlations between themselves and with spectral parameters (see
Wijnands \& van der Klis 1999; 
Psaltis, Belloni \& van der Klis 1999; Belloni, Psaltis \& van der Klis 2002; 
Markwardt, Swank \& Taam 1999; Vignarca et al. 2003).
This state includes the LS and HIMS: the latter is associated to small accretion
disk radii, faster jet ejection, steeper energy spectra and higher 
characteristic frequencies. In the HIMS, the corona component (see below)
starts contributing to the X-ray flux (Zdziarski et al. 2001; Rodriguez
et al. 2004).

In the soft states (HS and SIMS), the jet is suppressed (jet
radio and X-ray components are not observed). The flux is dominated by the
accretion disk component, which now has a higher temperature and a small
inner radius, possibly coincident with the innermost stable orbit around
the black hole, but an additional steep power-law component, with no evidence
of a high-energy cutoff up to $\sim$1 MeV is visible, which we associate
here (generically) to a corona. In the power spectra, no strong band-limited
noise component is observed, and transient QPOs of type A/B are observed,
with frequencies above a few Hz and not much variable.

\begin{figure}
  \includegraphics[height=.4\textheight]{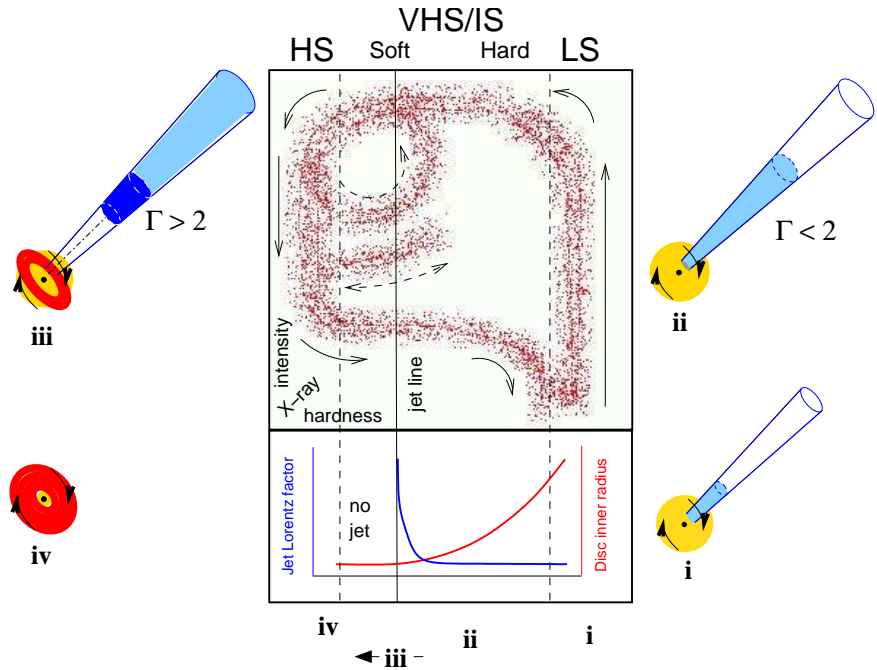}
  \caption{Schematic view of the accretion/ejection coupling in BHBs. The
  	top panel shows an idealized HID, the bottom panel shows the 
	$\Gamma$ factor of the jet and the value of the inner accretion 
	disk radius (see text). The pictures on the side sketch the 
	structure of the three components in the system: jet, 
	disk, and corona.
	From Fender, Belloni \& Gallo (2004).}
\end{figure}

This picture corresponds to the schematic outburst evolution shown 
in Fig. 3 (from Fender, Belloni \& Gallo 2004) .
In addition to the top horizontal branch in Figs. 1 and 3, transitions across
the jet line, which separates the two major states described above, can be
seen also on short time scales. In XTE J1859+226, very fast transitions 
are observed between type-B QPOs at 6 Hz and type-C QPOs at 8 Hz, i.e. between
the SIMS and the HIMS with high characteristic frequencies
(Casella et al. 2004). These transitions have been seen at times corresponding
to major radio activity (Casella, this volume). This association resembles
that seen in GRS 1915+105 (Klein-Wolt et al. 2002; Fender \& Belloni 2004).
These short-term transitions should be studied in more detail.
Although timing analysis of the fast variability of BHBs can
give us direct measurements of important parameters of the accretion flow, 
up to now we do not have unique models that permit this. Recent results show that
a clear association can be made between type-C QPO, strong band-limited noise
and the presence of a relativistic jet. In the framework of unifying models,
these results could play an important role.




\bibliographystyle{aipproc}   

\end{document}